\documentclass[preprint,12pt]{elsarticle}

\usepackage{amssymb}
\usepackage{amsthm}
\usepackage{amsmath}
\usepackage{bm}

\journal{Acta Astronautica}

\begin{document}

\begin{frontmatter}



\title{Adaptive Relative Orbit Control Considering Laser Ablation Uncertainty}


\author[KU]{Shun Isobe}
\author[KU]{Yasuhiro Yoshimura}
\author[KU]{Toshiya Hanada}
\author[JSAT]{Yuki Itaya}
\author[JSAT]{Tadanori Fukushima}

\affiliation[KU]{organization={Kyushu Univerisity},
            addressline={744 Motooka, Nishi-ku}, 
            city={Fukuoka},
            postcode={819-0395}, 
            country={Japan}}
\affiliation[JSAT]{organization={SKY Perfect JSAT Corporation},
            addressline={8-1 Akasaka 1-chome, Minato-ku}, 
            city={Tokyo},
            postcode={107-0052}, 
            country={Japan}}


\begin{abstract}
This study proposes a relative orbit control law for laser debris removal missions considering the uncertainties of laser ablation and atmospheric drag.  A removal spacecraft irradiates laser pulses to a target debris to generate the ablation force for deorbiting.  The deorbiting force lowers the target altitude, and the removal spacecraft must follow it to maintain its relative position for continuous laser irradiation. The difficulty stems from uncertainties of the magnitude of laser ablation and external disturbances such as atmospheric drag.  To tackle this problem, this study derives an adaptive control method using the Gaussian process regression to cancel the uncertainties with a nonparametric regression model.  Numerical simulations verify the proposed control law under the uncertainties of laser ablation and atmospheric drag.  The proposed control law can contribute to the realization of a safer and more secure mission not only for laser debris removal missions, but also for other on-orbit services.

\end{abstract}




\end{frontmatter}


\section{Introduction}
\label{sec:intro}
Orbital debris is a severe problem for sustainable space development.  To remediate the orbital environment, active debris removal (ADR) of at least five large debris per year is required~\cite{jc}.  Although several ADR methods have been proposed and experimented on orbit such as RemoveDEBRIS and ELSA-d~\cite{removeDebris,black}, these methods require direct contact operation.  On the other hand, an ADR method that uses laser ablation has an advantage in contactless operation~\cite{fukushima2021}.  Laser ablation is the process of ejecting a material in the vaporized state from a solid surface by irradiating intense laser pulses~\cite{tsuno2020a}.  The ablation force is generated as the reaction force of the gas ejected from the surface and is utilized to lower the orbital altitude of the target debris.  The laser ADR method has the following advantages~\cite{fukushima2021}: contactless debris removal and no need to carry fuel for deorbiting. 

In the laser ADR mission, a removal spacecraft irradiates laser pulses to a target debris to generate the ablation force for deorbiting.  The deorbiting force lowers the target altitude, and the removal spacecraft must follow the target to maintain the relative position for continuous laser irradiation.  Thus, formation keeping control in such powered flight is essential for the ADR mission using laser ablation.  The previous study has proposed a formation keeping control law for the laser ADR method~\cite{isb}.  However, the controller in~\cite{isb} is based on a feedforward design, and uncertainties of laser ablation and external disturbances, such as atmospheric drag, are not considered.  The uncertainty on the ablation thrust comes from the target's material and the spot size of the laser~\cite{tsuno2020a}.  Atmospheric drag is the primary disturbance force in low Earth orbit and has uncertainties in atmosphere models and ballistic coefficients.  The uncertainties of the disturbance forces may increase the relative position error, resulting in a high risk of collision.  For long-term debris removal missions, autonomous feedback control laws considering disturbances without manual adjustment are required to realize safer and more secure mission realization.  Therefore, this study derives the adaptive feedback relative orbit control considering both uncertainties.

This study derives a model reference adaptive control (MRAC) method using the Gaussian process (GP) regression that considers both uncertainties of laser ablation and atmospheric drag.  MRAC is one of the typical adaptive control methods and can automatically adjust the control system based on the characteristics of the observed real system.  Ulrich~\cite{ulrich} developed a relative orbit control law for elliptical orbits using MRAC.  Tiwari et al.~\cite{tiwari} developed the orbit and attitude MRAC and applied trajectory tracking near the asteroid under uncertainties of the mass and inertia of the spacecraft and the gravitational field.  Although conventional MRAC enables relative orbit control under uncertainties, the conventional controller requires parametric regression of the modeling error by an assumed basis function.  On the other hand, GP regression can learn nonparametric models without prior information~\cite{liu}. Thus, GP regression can cancel the modeling error in the MRAC architecture~\cite{chowdhary}. For formation keeping control in this study, an adaptive controller using GP has an advantage in controlling relative orbit without the assumption of uncertainties.

This paper is organized into the following sections.  In Section 2, this paper describes an overview of relative motion formulations and GP regression.  The Hill--Clohessy--Wiltshire equations of relative motion and the analytical solution for the design of the control law are derived.  In addition, an algorithm for GP regression using Bayes law and properties of Gaussian distributions is presented.  In Section 3, the relative orbit control law based on GP-MRAC is designed.  Specifically, the reference model, the control gains, and the control parameters are determined.  In the GP-MRAC architecture, the uncertainties of laser ablation and atmospheric drag are canceled by GP regression using observed information.  In Section 4, this paper verifies the performance of the proposed control law compared to the PD control law.  First, to verify that the proposed controller cancels the uncertainties, numerical simulations are performed under atmospheric drag disturbance.  
Second, considering the laser ADR method, numerical simulations are performed assuming formation, under uncertainty of the laser ablation thrust and atmospheric drag disturbance.  In actual operation, the relative orbit control using ON/OFF thrusters with constant magnitudes is considered.  That is, an algorithm is used that translates the continuous signal of GP-MRAC into the ON/OFF signal.

\section{Preliminaries}
\subsection{Relative Motion}
\begin{figure}[tb]
\centering
\includegraphics{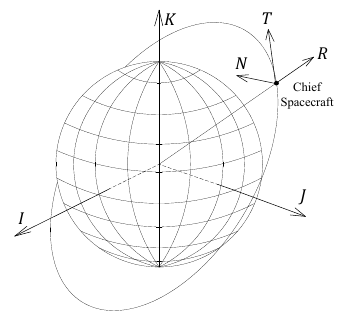}
\caption{Inertial frame and Hill frame.}
\label{fig:frame}
\end{figure}

This study uses two coordinate frames, as illustrated in Fig.~\ref{fig:frame}. The inertial frame has its origin at the Earth center.  The $\bm{I}$ axis is along the vernal equinox direction, the $\bm{K}$ axis is along the rotational axis of the Earth, and the $\bm{J}$ axis completes the right-handed frame.  The relative motion of a deputy spacecraft with respect to a chief spacecraft is described with the Hill frame whose origin is at the position of the chief spacecraft.  The $\bm{R}$ axis is aligned with the outward radial direction, the $\bm{N}$ axis is parallel to the orbital angular momentum vector of the chief spacecraft, and the $\bm{T}$ axis completes the right-handed frame.

Consider two spacecraft orbiting Earth.  Let $\bm{\rho}$ be the relative position of the deputy spacecraft with respect to the chief spacecraft in the Hill coordinates as $\bm{\rho} =x\bm{R}+ y \bm{T} + z\bm{N}$.
Since the Hill frame is the rotating frame, the equations of relative motion in the Hill frame are given by 
\begin{align}
\ddot{x} - 2n\dot{y}-3n^2x &= u_x + \Delta_x \label{eq:HCWeqX}\\
\ddot{y}+2n^2\dot{x} &= u_y+\Delta_y \label{eq:HCWeqY}\\
\ddot{z}+n^2z &= u_z+\Delta_z \label{eq:HCWeqZ}
\end{align}
where $n$ is the mean motion of the chief, $\bm{u}=[u_{x},u_{y},u_{z}]^{T}$ is the control thrust vector, and $\bm{\Delta} = [\Delta_{x},\Delta_{y},\Delta_{z}]^{T}$ is the disturbance acceleration in the Hill frame.
These equations are called the Hill--Clohessy--Wiltshire (HCW) equations of relative motion.

It is convenient to rewrite the HCW equations in state-space form.  Defining the state vector as $\bm{x}=[x,y,z,\dot{x},\dot{y},\dot{z}]^{T}$, Eqs.~\eqref{eq:HCWeqX},~\eqref{eq:HCWeqY}, and~\eqref{eq:HCWeqZ} are written as
\begin{align}
\dot{\bm{x}} = A\bm{x} + B(\bm{u}+ \bm{\Delta}) \label{eq:HCWmatrix}
\end{align}
where
\begin{align}
A &= \begin{bmatrix} 0 & 0 & 0 & 1 & 0 & 0\\  0 & 0 & 0 & 0 & 1 & 0\\  0 & 0 & 0 & 0 & 0 & 1\\ 3 n^2 & 0 & 0 & 0 & 2 n & 0\\  0 & 0 & 0 & - 2 n & 0 & 0\\  0 & 0 & - n^2 & 0 & 0 & 0 \end{bmatrix}\\
B &= \begin{bmatrix} 0 & 0 & 0\\ 0 & 0 & 0\\ 0 & 0 & 0\\ 1 & 0 & 0\\ 0 & 1 & 0\\ 0 & 0 & 1 \end{bmatrix} 
\end{align}
The discrete time system of Eq.~\eqref{eq:HCWmatrix} is obtained using the variation of parameters method as follows. (See~\ref{Appendix:matrix} for detailed matrices $\Phi$ and $\Psi$). 
\begin{align}
\bm{x}(t) =& e^{A(t-t_0)}\bm{x}(t_0) + \int_{t_0}^t e^{A(\tau-t_0)}Bd\tau (\bm{u}+ \bm{\Delta}) \\
=& \Phi \bm{x}(t_0) + \Psi (\bm{u}+ \bm{\Delta}) \label{eq:HCW_dis}
\end{align}
Assuming $\bm{u}=\bm{\Delta}=\bm{0}$, the analytical solution of Eq.~\eqref{eq:HCWmatrix} is written as
\begin{align}
x&=\sqrt{\left(3x_0+\frac{2{\dot{y}}_0}{n}\right)^2+\left(\frac{{\dot{x}}_0}{n}\right)^2}\cos\left(nt+\alpha\right)+4x_0+\frac{2{\dot{y}}_0}{n} \\
y&=2\sqrt{\left(3x_0+\frac{2{\dot{y}}_0}{n}\right)^2+\left(\frac{{\dot{x}}_0}{n}\right)^2}\sin\left(nt+\alpha\right)+\left(6nx_0+3\dot{y_0}\right)t \nonumber\\
&{+y}_0-\frac{2\dot{x}_0}{n} \label{eq:analySolY}\\
z&=\sqrt{\left(z_0\right)^2+\left(\frac{{\dot{z}}_0}{n}\right)^2}\sin\left(nt+\beta\right)
\end{align}
where the subscript ``0'' represents the values at $t=t_{0}$. The phase angle $\alpha$ and $\beta$ can be expressed as 
\begin{align}
\alpha &=\tan^{-1}{\left(\frac{{\dot{x}}_0}{3{nx}_0+2{\dot{y}}_0}\right)} \\
\beta &=\tan^{-1}{\left(\frac{nz_0}{{\dot{z}}_0}\right)}
\end{align}
From the analytical solution in Eq.~\eqref{eq:analySolY}, the bounded relative orbit is obtained by choosing the initial condition as
\begin{align}
6nx_{0} + 3\dot{y}_{0} = 0
\end{align}
In the in-plane motion, the bounded relative orbit is an ellipse where the semi-major axis in the along-track direction is twice as large as the semi-minor axis in the radial direction. In this study, the bounded relative orbit is set to the desired relative orbit.

\subsection{Differential drag and uncertainties}
The differential drag is relative accelerations generated by differentials in the atmospheric drag between chief and deputy and is given by
\begin{align}
\bm{\Delta}_{\rm drag} = \frac{1}{2}\frac{C_{D,d}A_{d}}{M_{d}}\rho_{d} \|\bm{v}_{{\rm rel},d}\|\bm{v}_{{\rm rel},d} - \frac{1}{2}\frac{C_{D}A_{c}}{M_{c}}\rho_{c} \|\bm{v}_{{\rm rel},c}\|\bm{v}_{{\rm rel},c}
\end{align}
where $A$ is the cross-sectional area, $M$ is the mass, $C_{D}$ is the aerodynamic drag coefficient, $\rho$ is the atmospheric density, and $\bm{v}_{\rm rel}$ is the spacecraft velocity relative to the atmosphere.  The variables with the subscript ``$d$'' refer to those of the deputy spacecraft, and the variables with the subscript ``$c$'' refer to those of the chief spacecraft.  
Differential drag has mainly uncertainties in atmospheric density and drag coefficient. 
Moreover, the cross-sectional area and mass need to be estimated for an uncooperative target, which also increases uncertainties.

\subsection{Model Reference Adaptive Control}
\begin{figure}[tb]
\centering
\includegraphics[width=8cm]{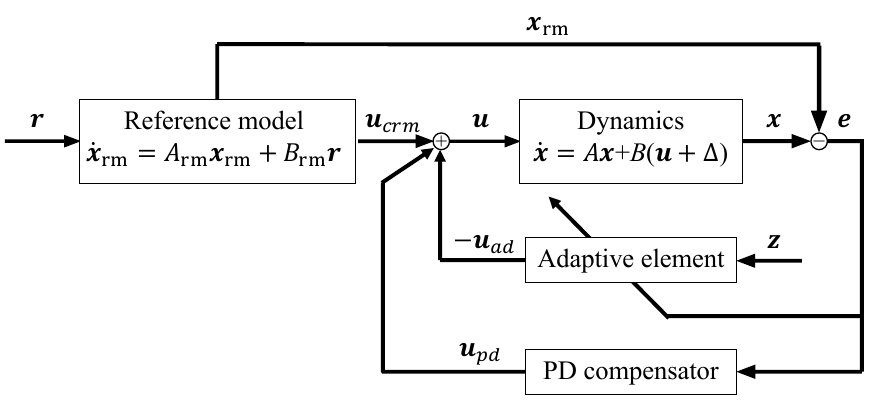}
\caption{Model reference adaptive control flow.}
\label{fig:MRAC}
\end{figure}

The MRAC can automatically adjust the control system based on the characteristics of the real system observed.  
In MRAC, the desired response of a real system is designed by a reference model, and the controller can adjust so that the real system corresponds to the reference model.  In other words, this control law converges the tracking error $\bm{e}=\bm{x}-\bm{x}_{\rm rm}$ between the actual state $\bm{x}$ and the reference state $\bm{x}_{\rm rm}$ to zero.  Figure~\ref{fig:MRAC} shows the control flow of the MRAC.  The reference model that characterizes the desired response can be designed as follows.
\begin{align}
\dot{\bm{x}}_{\rm rm} = A_{\rm rm}\bm{x}_{\rm rm} + B_{\rm rm}\bm{r}  \label{eq:refModel}
\end{align}
where $A_{\rm rm}$ is the reference state matrix, $B_{\rm rm}$ is the reference control matrix, and $\bm{r}$ is the reference signal.  
The reference signal is determined so that the reference state $\bm{x}_{\rm rm}$ is driven to the desired state.

The MRAC law consists of a linear feedback part $\bm{u}_{\rm pd}$, a linear feedforward part $\bm{u}_{\rm crm}$, and an adaptive part $\bm{u}_{\rm ad}$ as
\begin{align}
\bm{u} &= \bm{u}_{\rm pd} + \bm{u}_{\rm crm} - \bm{u}_{\rm ad}  \label{eq:inputU}
\end{align}
where
\begin{align}
\bm{u}_{\rm pd} &= -K_{\rm pd} \bm{e}  \label{eq:uPD} \\
\bm{u}_{\rm crm} &= \begin{bmatrix}
    K_{m} &  K_{r}
\end{bmatrix} \begin{bmatrix}
     \bm{x}_{\rm rm}  \\
     \bm{r} 
\end{bmatrix} \label{eq:uCRM}
\end{align}
In Eqs.~\eqref{eq:uPD} and~\eqref{eq:uCRM}, $K_{\rm pd}$, $K_{m}$, and $K_{r}$ are the control gain matrices.  
Using Eqs.~\eqref{eq:HCWmatrix},~\eqref{eq:inputU}--\eqref{eq:uCRM}, the dynamics of tracking error can be written as
\begin{align}
\dot{\bm{e}} &= (A-BK_{\rm pd})\bm{e} + (A - A_{\rm rm}) \bm{x}_{\rm rm} - B_{\rm rm} \bm{r} + \bm{u}_{\rm crm} + B(\bm{\Delta} - \bm{u}_{\rm ad}) \\
&= (A-BK_{\rm pd})\bm{e} + (A - A_{\rm rm} + B K_{m}) \bm{x}_{\rm rm} \nonumber \\
& + (BK_{r} - B_{\rm rm})\bm{r} + B(\bm{\Delta} - \bm{u}_{\rm ad}) \label{eq:dotE}
\end{align}
To converge the tracking error $\bm{e}$, the MRAC law selects the control gain as follows.
\begin{align}
A + B K_{m} &= A_{\rm rm} \label{eq:AKr}\\
BK_{r} &= B_{\rm rm} \label{eq:BKr}
\end{align}
When the actual system $A$ and $B$ are unknown with large uncertainty, the MIT rule is used to determine the gains $K_{m}$ and $K_{r}$~\cite{aastrom2013adaptive}.  
Using Eqs.~\eqref{eq:dotE}--\eqref{eq:BKr}, the dynamics of the tracking error is obtained as
\begin{align}
\dot{\bm{e}} = (A - B K_{\rm pd})\bm{e} + B(\bm{\Delta} - \bm{u}_{\rm ad})
\end{align}
The control gain matrix $K_{\rm pd}$ is chosen so that $A-BK_{\rm pd}$ becomes a Hurwitz matrix.  

The adaptive part $\bm{u}_{\rm ad}$ should be designed to cancel the uncertain disturbance $\bm{\Delta}(\bm{z})$, where $\bm{z}$ is a parameter of the disturbance $\bm{\Delta}(\bm{z})$.  
For example, in the case of atmospheric drag, the parameter $\bm{z}$ is assumed to include atmospheric density, orbital velocity, ballistic coefficient, and latitude argument.

\subsection{Gaussian Process Regression}
GP regression is a data-driven machine learning model without prior information and is suitable for regression problems.  In this study, the uncertainties of laser ablation and atmospheric drag are canceled by GP regression using observed information.  A GP is defined as a collection of random variables and has a Gaussian distribution with mean $m(\bm{z})$ and covariance $k(\bm{z}_{n},\bm{z}_{n'})$ where $\bm{z},\bm{z}_{n}$, and $\bm{z}_{n'}$ are input state vectors.
The output of the GP regression is represented as
\begin{align}
y\sim\mathcal{GP}\left(m\left(\bm{z}\right),k\left(\bm{z}_{n},\bm{z}_{n'}\right)\right)
\end{align}
Given some set of noisy data points $\bm{y}=[y_{1},y_{2},\dots, y_{i}]^{T}$ at corresponding input set $\bm{Z}=[\bm{z}_{1},\bm{z}_{2},\dots,\bm{z}_{i}]^{T}$, the GP predicts the expected value ${\hat{y}}_{i+1}$ at a new input $\bm{z}_{i+1}$.  
Using the Bayesian law, the mean $m(\bm{z}_{i+1})$ and the covariance $\Sigma(\bm{z}_{i+1})$ can be calculated from the training data $Z$ and $\bm{y}$ as follows.
\begin{align}
m(\bm{z}_{i+1}) &= \bm{\alpha}^{T} \bm{k}(Z, \bm{z}_{i+1})\\
\Sigma(\bm{z}_{i+1}) &= k(\bm{z}_{i+1}, \bm{z}_{i+1}) - \bm{k}^{T} (Z,\bm{z}_{i+1})C\bm{k}(Z,\bm{z}_{i+1})
\end{align}
where 
\begin{align}
C &= \left[K(Z,Z) + \omega^{2}I\right]^{-1} \label{eq:largeC}\\
\bm{\alpha} &=C\bm{y}
\end{align}
In Eq.~\eqref{eq:largeC}, $\omega^{2}$ is the variance of the observation noise that follows Gaussian distribution.
The covariance of the input vectors is calculated with a kernel function, and the Gaussian kernel in Eq.~\eqref{eq:gKernel} is commonly selected.
\begin{align}
k(\bm{z}_{n},\bm{z}_{n'}) &= \theta_{1}\exp{\left(\frac{\|\bm{z}_{n}-\bm{z}_{n'}\|^{2}}{\theta_{2}}\right)}\label{eq:gKernel}
\end{align}
The kernel matrix $K$ consists of the kernel function, e.g., $K_{j,k} = k(\bm{z}_{j},\bm{z}_{k})$.
In the kernel function, $\theta_{1}$ and $\theta_{2}$ are hyperparameters.
For example, a large $\theta_{2}$ results in a smooth function approximation, whereas a small $\theta_{2}$ can capture the details, but will also be more sensitive to noise.  In this study, the hyperparameters are determined empirically, but can be selected by optimization.

\section{Adaptive Relative Orbit Control}
\subsection{Gaussian Process Model Reference Adaptive Control}
In MRAC, the adaptive part $\bm{u}_{\rm ad}$ using a radial basis function (RBF) neural network (NN) is one of the typical MRAC methods.  However, this RBF method is the parametric regression of the disturbance by an assumed basic function. Thus, this study uses GP regression for the adaptive part because GP can learn nonparametric models.

The GP-MRAC uses the disturbance force $\bm{\Delta}(\bm{z})$ as observation data and updates the training data $Z$ and $\bm{y}$ automatically.  
When measurements $\bm{\Delta}'=\bm{\Delta}(\bm{z})+\bm{\epsilon}$ are corrupted by Gaussian white noise $\bm{\epsilon}$, the resulting output also follows a Gaussian distribution.  
This results in a model of the stochastic process $\bm{\Delta}(\bm{z})$ as follows.
\begin{align}
\Delta_{j} \sim \mathcal{GP}(m_{j}(\bm{z}), k(\bm{z}_{n}, \bm{z}_{n'})) \hspace{10pt} (j=x,y,z)
\end{align}
The adaptive part $\bm{u}_{\rm ad}$ is set to be equal to the estimate of the mean as follows.
\begin{align}
\bm{u}_{\rm ad}(\bm{z}) = [m_{x}(\bm{z}), m_{y}(\bm{z}), m_{z}(\bm{z})]^{T}
\end{align}
To store training data $Z$ and $\bm{y}$ on board, an online method is needed to restrict the number of data, and this paper implements the sparse GP algorithm~\cite{csato2002sparse,csato2000sparse}.  
This algorithm approximates the Kullback--Leibler (KL) divergence between the current GP and the latest GP adding new measurements, and selects the data point to be deleted with the largest KL divergence case.

\subsection{Relative Orbit Control Law Based on GP-MRAC}
This subsection designs the relative orbit control law based on GP-MRAC. First, the reference model is designed. In this study, the reference model is set to match the ideal real system without disturbance described as
\begin{align}
A_{\rm rm} &= A \\
B_{\rm rm} &= B \\
r &= -K_{\rm pd} (\bm{x}_{\rm rm} - \bm{x}_{d})
\end{align}
where $\bm{x}_{d}$ is the desired state vector.

For the relative orbit control of a deputy spacecraft with respect to a chief spacecraft in a circular orbit, the control gain $K_{\rm pd}$ is selected as follows~\cite{alfriend2009}.
\begin{align}
K_{\rm pd} = \begin{bmatrix}
    4n^{2} & 0 & 0 & cn & 2n & 0\\
    0 & n^{2} & 0 & -2n & cn & 0 \\
    0 & 0 & n^{2} & 0 & 0 & cn
\end{bmatrix} \label{eq:Kpd}
\end{align}
where $c$ is a positive constant, which can be selected to achieve a desired convergence speed.  
In this study, the constant $c$ is set to $c=1.0$. 
Using Eq.~\eqref{eq:Kpd}, $A-BK_{\rm pd}$ in Eq.~\eqref{eq:dotE} can be rewritten as follows.
\begin{align}
A-BK_{\rm pd} = \begin{bmatrix}
     &  &  &  &  &  \\
     & 0_{3\times 3} & & & I_{3} & \\
    & & & & &  \\
    -n^{2} & 0 & 0 & -cn & 0 & 0 \\
    0 & -n^{2} & 0 & 0 & -cn & 0\\
    0 & 0 & -2n^{2} & 0 & 0 & -cn
\end{bmatrix}
\end{align}
where $0_{3\times 3}$ is the zero matrix and $I_{3}$ is the identity matrix.
Thus, the matrix $A-BK_{\rm pd}$ is the Hurwitz matrix. 
The control gains $K_{m}$ and $K_{r}$ of the feedforward part $\bm{u}_{\rm crm}$ are determined by Eqs.~\eqref{eq:AKr} and~\eqref{eq:BKr}.  
In this study, the control gains are determined as follows.
\begin{align}
K_{m} &= 0_{6\times 6} \\
K_{r} &= I_{6}
\end{align}
Finally, an adaptive part $\bm{u}_{\rm ad}$ using GP is designed. 
The input parameter $\bm{z}$ in the GP should be selected taking into account the characteristic of uncertain disturbance acceleration $\bm{\Delta}(\bm{z})$, which will be discussed in Section~\ref{sec:NS}.

GP-MRAC requires the state derivative $\dot{\bm{x}}$ to update the training data $Z$ and $\bm{y}$. Obtaining high-frequent information on acceleration is not preferable due to noisy measurements or lack of sensors.  In this study, the following calculation from Eq.~\eqref{eq:HCW_dis} is used to obtain the disturbance acceleration $\bm{\Delta}'_{i}$ from the state vector at time $t_i$ and $t_{i+1}$.
\begin{align}
\label{eq:delta0}
\bm{\Delta}'_{i}  &= (\Psi^T \Psi)^{-1} \Psi^T (\bm{x}(t_{i+1}) - \Phi \bm{x}(t_{i})) - \bm{u}(t_{i}) + \bm{\epsilon}\\
\label{eq:z}
\bm{z}_{i} &= \frac{\bm{z}(t_{i+1}) + \bm{z}(t_{i})}{2}
\end{align}

\section{Numerical Simulation}\label{sec:NS}
\subsection{Simulation Condition}
Numerical simulations are performed for two test cases to verify the proposed controller compared to the PD controller under uncertainties.  The first case considers the uncertainty of atmospheric drag, and the second case considers the uncertainty of laser ablation thrust.
Furthermore, from the point of view of actual operation, in the second case the relative orbit control using ON/OFF thrusters with constant magnitude is considered.
The simulation conditions are summarized in Table~\ref{tab:condition}.  It is noted that this study considers the target debris as the chief and the removal spacecraft as the deputy.  
\begin{table*}[tb]
\begin{center}
\caption{Simulation condition} \label{tab:condition}
\begin{tabular}{lcc} 
\hline \hline
Epoch & & 2015/01/01 \\ \hline
Chief & & Target debris \\
& & Circular orbit at altitude 450 km \\
& Shape & Box-wing \\
& Mass & 100 kg \\
& Area to mass ratio & $0.045~{\rm m^{2}/kg}$ \\ \hline
Deputy & & Removal spacecraft\\
& Shape & Box \\
& Mass & 150 kg \\
& Area to mass ratio & $0.004~{\rm m^{2}/kg}$ \\ \hline
Atmosphere model & & Jaccia--Bowman 2008~\cite{bowman2008new} \\
\hline \hline
\end{tabular}
\end{center}
\end{table*}

\subsection{Uncertainty of Differential Drag}
\begin{figure}[tb]
\centering
\includegraphics[width=8cm]{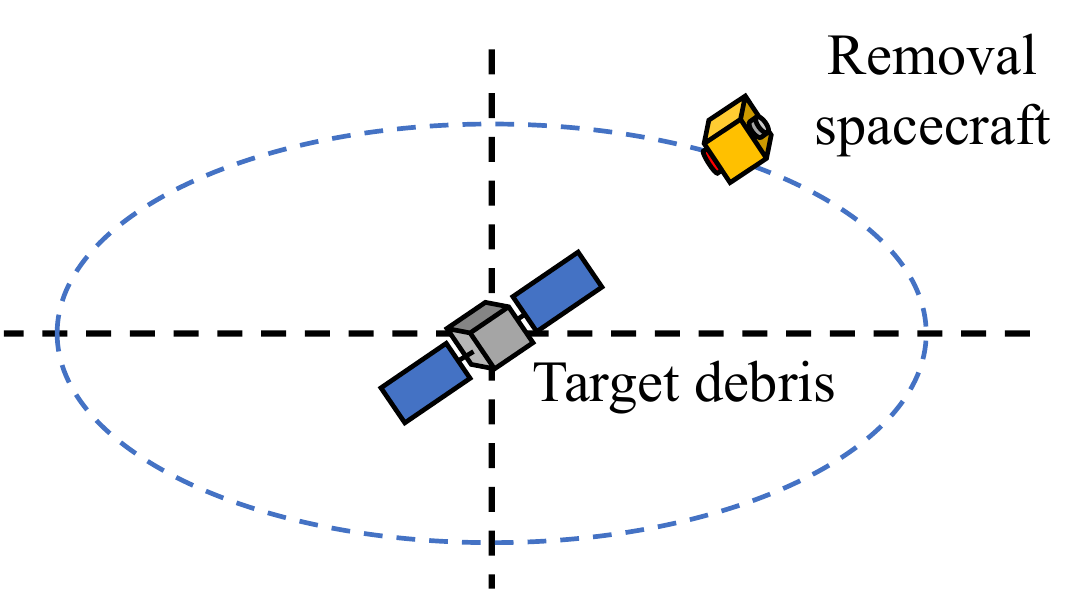}
\caption{Desired relative orbit.}
\label{fig:relMotion_SE}
\end{figure}

This case considers the uncertainty of atmospheric drag.  
The proposed controller learns from training data and predicts the differential drag including uncertainties.  The mean argument of latitude $u$ and the sun phase angle $\phi_s$ are used for the input parameter $\bm{z}$ in the GP regression, because the atmospheric drag is periodic and correlated with the positions of the Sun and Earth.  
Thus, the kernel function is written as follows.
\begin{align}
\label{eq:k_propose}
k(\bm{z},\bm{z}') &= \exp\left( - \frac{ \sin^2( (u - u') / 2)}{2 \sigma_u^2} - \frac{ \sin^2( (\phi_s-\phi_s') / 2) }{2 \sigma_\phi^2} \right)
\end{align}
where the hyperparameters $\sigma_u$ and $\sigma_\phi$ are set to $\sigma_u = 0.25$ and $\sigma_\phi= 0.70$, respectively.
The desired relative orbit is an ellipse of the semi-major axis 60~m in the along-track direction and the semi-minor axis 30 m in the radial direction as shown in Fig.~\ref{fig:relMotion_SE}.
This simulation case assumes continuous thrust without uncertainties.

\begin{figure}[tb]
\centering
\includegraphics[width=8cm]{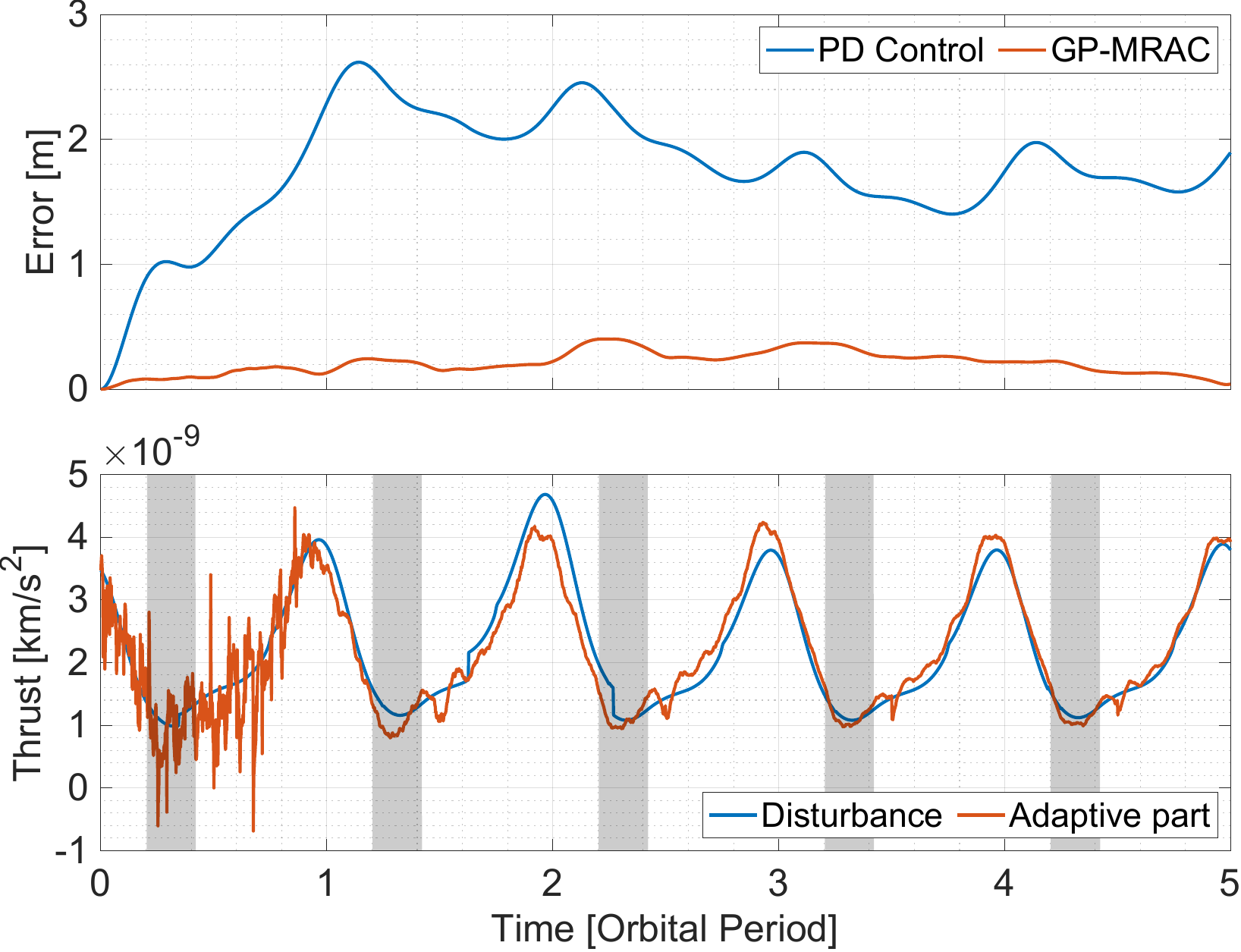}
\caption{Tracking error and thrust.}
\label{fig:case1}
\end{figure}

Figure~\ref{fig:case1} describes the time history of the tracking error and the adaptive part of the thrust.  
The shaded area indicates that the spacecraft is in the shadow area.  The tracking error shows that both the PD control (blue) and GP-MRAC (red) maintain the desired relative orbit, but the PD control has steady deviation.  In the adaptive part of the thrust, the GP regression (red) shows better performance to cancel the  actual differential drag (blue), though it overshoots at peak points.  Thus, the GP-MRAC excels the PD controller in terms of tracking error because the adaptive part successfully cancels the differential drag.

\subsection{Uncertainty of Laser ablation force}
This case considers the uncertainty of the laser ablation thrust.  
In the laser ADR method, the removal spacecraft irradiates laser pulses to the target debris to generate the ablation force for deorbiting.  
The deorbiting force lowers the target, and the removal spacecraft must follow the target debris to maintain its relative position for continuous laser irradiation.  
For simplicity, the ablation force is modeled by trigonometric functions as follows.
\begin{align}
\bm{\Delta}_{\rm ab} = a_{0} \frac{\bm{F}_{\rm ab}}{M} + a_{1}\frac{\bm{F}_{\rm ab}}{M}\cos{u}+ a_{2}\frac{\bm{F}_{\rm ab}}{M}\sin{u}
\end{align}
where $\bm{F}_{\rm ab}$ is the laser ablation force without uncertainty.  Taking into account the irradiation of the laser pulse to the aluminum surface of the target debris, $\bm{F}_{\rm ab}$ is set to $\bm{F}_{\rm ab}=[0, -0.72, 0]^{T}$~mN~\cite{tsuno2020a}. 
Arbitrary weights $a_{0},a_{1}$, and $a_{2}$ are set to 1.0, 0.10, and 0.10, respectively.  

\begin{figure}[tb]
\centering
\includegraphics[width=8cm]{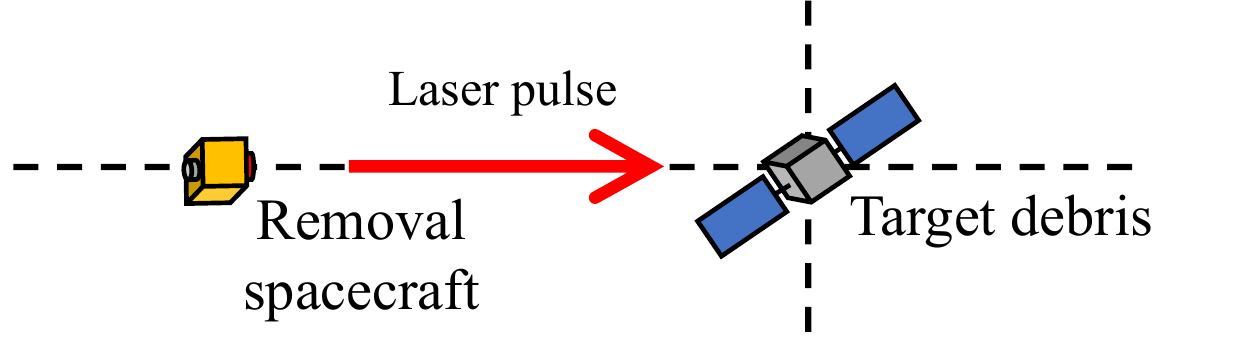}
\caption{Desired relative orbit for continuous laser irradiation.}
\label{fig:relMotion}
\end{figure}

The desired relative orbit is stationary keeping 100~m in front of the target debris in the along-track direction as shown in Fig.~\ref{fig:relMotion}.  In actual operation, the relative orbit control using ON/OFF thrusters with constant magnitudes is considered.  
Therefore, this study adopts the following modulation method that translates the continuous signal of GP-MRAC $\bm{u}$ into an ON/OFF signal with constant thrust $F$ ~\cite{ono2021gnc}.
\begin{align}
    ||\bm{u}|| T_{\rm int} =& F \Delta t \label{eq:timeInt}\\
    \Delta t =& T_{\rm int} \frac{||\bm{u}||}{F}
\end{align}
where $T_{\rm int}$ is time intervals of control, and $\Delta t$ is ON/OFF thruster injection duration.  In this method, the $F\bm{u}/||\bm{u}||$ maneuver is performed for $\Delta t$ seconds.  
Assuming that a cold gas thruster is used, $F$ is set to 0.1 N and the time intervals of control $T_{\rm int}$ is set to 300 s.

\begin{figure}[tb]
\centering
\includegraphics[width=8cm]{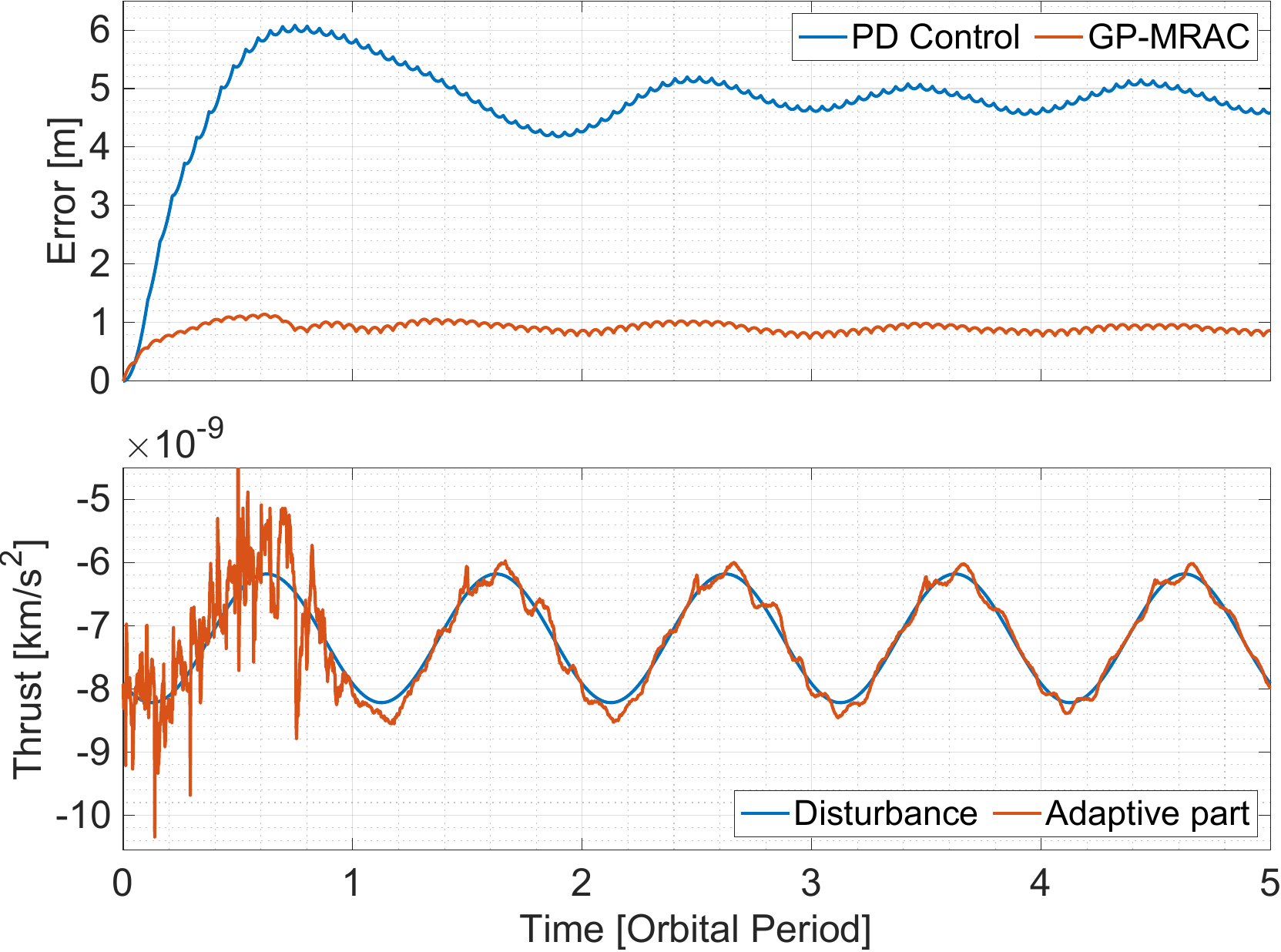}
\caption{Tracking error and thrust (Case 2).}
\label{fig:case2}
\end{figure}

\begin{figure}[tb]
\centering
\includegraphics[width=8cm]{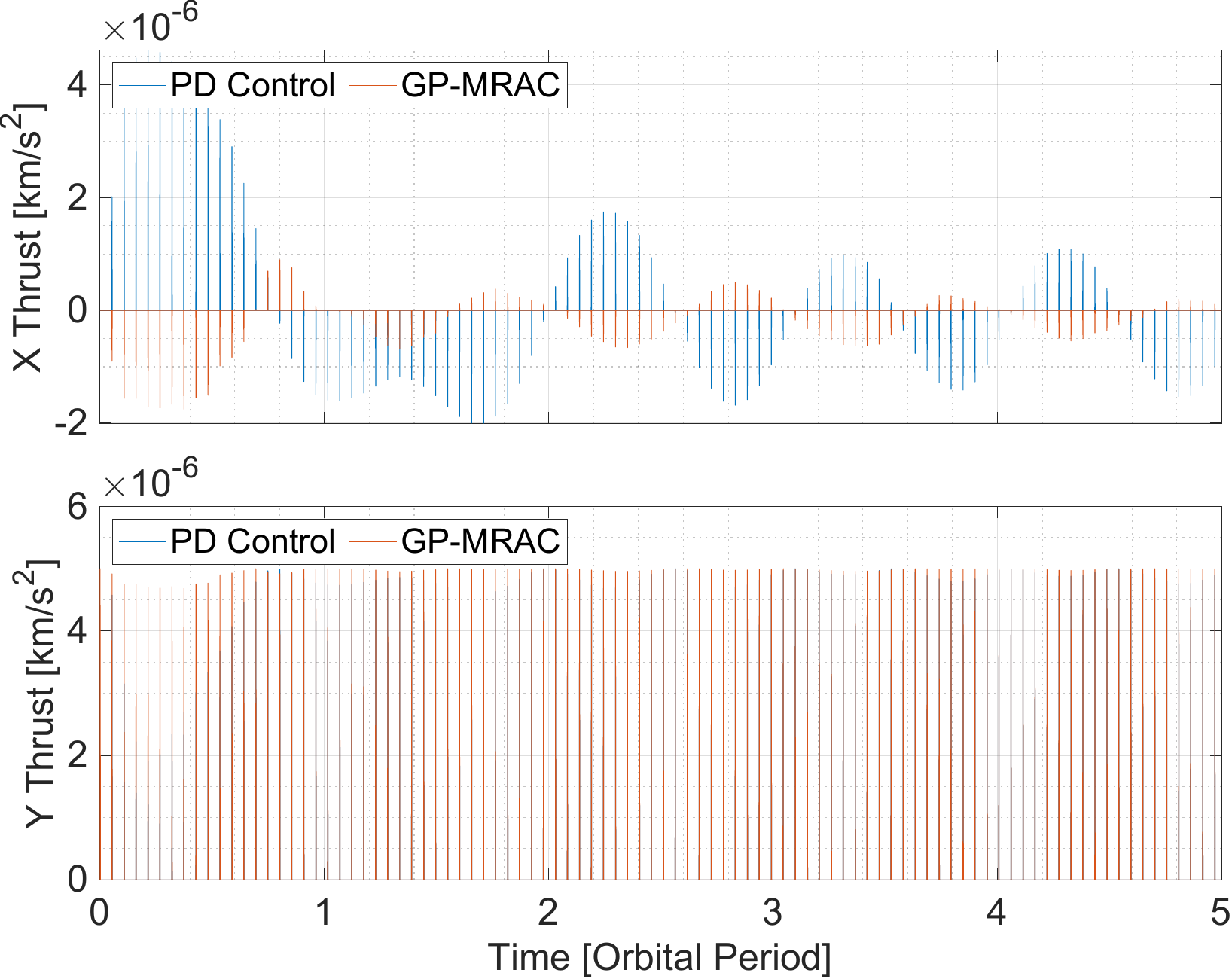}
\caption{PWPF modulated pulses}
\label{fig:thrust}
\end{figure}

Figure~\ref{fig:case2} describes the time histories of the tracking error and the adaptive part of the thrust.  
Both the PD control and the GP-MRAC have the steady deviation, although the GP regression (red) shows a better performance to cancel the actual differential drag and laser ablation thrust (blue).  
Figure~\ref{fig:thrust} shows the time history of the maneuver, which confirms the transformation from the continuous signal of GP-MRAC to the ON/OFF signal.
The maneuvers are frequently implemented in the along-track direction to compensate the relative position error caused by the uncertain ablation thrust.
This result shows that the time interval method in Eq.~\eqref{eq:timeInt} is suitable for the GP-MRAC.

\section{Conclusions}
This study dealt with the adaptive relative orbit control law in laser debris removal missions considering the uncertainties of laser ablation and atmospheric drag. This study derived the relative orbit controller based on the Gaussian process model reference adaptive control (GP-MRAC), which can automatically adjust the control system based on the characteristics of the observations and realize the long-term mission.  The uncertain disturbance acceleration is canceled by GP regression using observed information.  
Numerical simulations showed that GP-MRAC has a better performance in terms of tracking error compared to the PD controller, because the adaptive part of GP-MRAC successfully cancels the differential drag.  In addition, this study revealed that effective regression can be realized by selecting the mean argument of latitude and the sun phase angle as training data in GP because the atmospheric drag is periodic and correlated with the position of the Sun and the Earth.



\appendix
\section{}
\label{Appendix:matrix}
The elements of the matrix $\Phi$ in Eq.~\eqref{eq:HCW_dis} are
\begin{align}
\Phi &= \left[\begin{array}{cc} \Phi_{11} &\Phi_{12} \\ \Phi_{21} &\Phi_{22} \end{array}\right] \\
\Phi_{11} &= \left[\begin{array}{ccc} 4-3 c \left(n \left(t-t_{0}\right)\right) & 0 & 0\\ 6 s\left(n \left(t-t_{0}\right)\right) - 6n (t-t_{0}) & 1 & 0\\ 0 & 0 & c\left(n\left(t-t_{0}\right)\right) \end{array}\right]\\
\Phi_{12} &= \frac{1}{n}\left[\begin{array}{ccc} s\left(n\left(t-t_{0}\right)\right) & -2\left(c\left(n \left(t-t_{0}\right)\right)-1\right) & 0\\ 2\left(c\left(n \left(t-t_{0}\right)\right)-1\right) & 4s\left(n \left(t-t_{0}\right)\right)-3(t-t_{0}) & 0\\ 0 & 0 & s\left(n \left(t-t_{0}\right)\right) \end{array}\right]\\
\Phi_{21} &= n \left[\begin{array}{ccc} 3 s\left(n \left(t-t_{0}\right)\right) & 0 & 0\\ 6\left(c\left(n \left(t-t_{0}\right)\right)-1\right) & 0 & 0\\ 0 & 0 & -s\left(n \left(t-t_{0}\right)\right) \end{array}\right]\\
\Phi_{22} &= \left[\begin{array}{ccc} c\left(n \left(t-t_{0}\right)\right) & 2 s\left(n \left(t-t_{0}\right)\right) & 0\\ -2 s\left(n \left(t-t_{0}\right)\right) & 4 c\left(n \left(t-t_{0}\right)\right)-3 & 0\\ 0 & 0 & c\left(n \left(t-t_{0}\right)\right) \end{array}\right]
\end{align}
where $s_{(\cdot)}$ and $c_{(\cdot)}$ denote $\sin (\cdot)$ and $\cos (\cdot)$, respectively. 
The elements of the matrix $\Psi$ in Eq.~\eqref{eq:HCW_dis} are
\begin{align}
\Psi &= \left[\begin{array}{c} \Psi_{1}\\ \Psi_{2} \end{array}\right] \\
\Psi_{1} &= \frac{1}{n^2}\left[\begin{array}{ccc} 2 s^2\left(\frac{n \left(t-t_{0}\right)}{2}\right) & -2 s\left(n \left(t-t_{0}\right)\right)+2n \left(t-t_{0}\right) & 0\\ 2 s\left(n \left(t-t_{0}\right)\right)-2 n \left(t-t_{0}\right) &8s^2\left(\frac{n \left(t-t_{0}\right)}{2}\right)-\frac{3}{2} n^2(t-t_{0})^2& 0\\ 0 & 0 & 2 s^2\left(\frac{n \left(t-t_{0}\right)}{2}\right) \end{array}\right]\\
\Psi_{2} &= \frac{1}{n} \left[\begin{array}{ccc} s\left(n \left(t-t_{0}\right)\right) & 4 {s^2\left(\frac{n \left(t-t_{0}\right)}{2}\right)} & 0\\ -4s^2\left(\frac{n \left(t-t_{0}\right)}{2}\right) & -3n(t-t_{0})+4 s\left(n \left(t-t_{0}\right)\right) & 0\\ 0 & 0 & s\left(n \left(t-t_{0}\right)\right) \end{array}\right]
\end{align}

\bibliographystyle{elsarticle-num} 
\bibliography{laserAdaptive}

@Article{jc,
author = {Liou, J.-C.}, 
title = {An active debris removal parametric study for LEO environment remediation}, 
journal = {Advances in Space Research}, 
volume = {47}, 
number = {11}, 
pages = {1865–1876}, 
year = {2011}, }

@Article{removeDebris,
author = {Forshaw, Jason L. and Aglietti, Guglielmo S. and Navarathinam, Nimal and Kadhem, Haval and Salmon, Thierry and Pisseloup, Aurélien and Joffre, Eric and Chabot, Thomas and Retat, Ingo and Axthelm, Robert and Barraclough, Simon and Ratcliffe, Andrew and Bernal, Cesar and Chaumette, François and Pollini, Alexandre and Steyn, Willem H.}, 
title = {RemoveDEBRIS: An in-orbit active debris removal demonstration mission}, 
journal = {Acta Astronautica}, 
volume = {127}, 
pages = {448–463}, 
year = {2016}, }

@Proceedings{black,
author = {Blackerby, Chris and Okamoto, Akira and Iizuka, Seita and Kobayashi, Yusuke and Fujimoto, Kohei and Seto, Yuki and Fujita, Sho and Iwai, Takashi and Okada, Nobu and Forshaw, Jason}, 
title = {The ELSA-d end-of-life debris removal mission: preparing for launch}, 
booktitle = {The ELSA-d end-of-life debris removal mission: preparing for launch}, 
volume = {Proceedings of the International Astronautical Congress, IAC 8}, 
year = {2019}, }

@Proceedings{fukushima2021,
author = {Fukushima, Tadanori and Hirata, Daisuke and Adachi, Kazuma and Itaya, Yuki and Yamada, Jun and Tsuno, K and Ogawa, T and Saito, N and Sakashita, M and Wada, S}, 
title = {End of Life Deorbit Service with a Pulsed Laser Onboard a Small Satellite}, 
booktitle = {End of Life Deorbit Service with a Pulsed Laser Onboard a Small Satellite}, 
volume = {Proceedings of the 8th European Conference on Space Debris}, 
year = {2021}, }

@Article{tsuno2020a,
author = {Tsuno, K and Wada, S and Ogawa, T and Ebisuzaki, T and Fukushima, T and Hirata, D and Yamada, J and Itaya, Y}, 
title = {Impulse measurement of laser induced ablation in a vacuum.}, 
journal = {Opt Express}, 
volume = {28}, 
number = {18}, 
pages = {25723–25729}, 
year = {2020}, }

@Proceedings{isb,
author = {Isobe, Shun and Yoshimura, Yasuhiro and Hanada, Toshiya and Itaya, Yuki and Fukushima, Tadanori}, 
title = {Formation Keeping Control For Simultaneous Deorbit Using Laser Ablation}, 
booktitle = {Formation Keeping Control For Simultaneous Deorbit Using Laser Ablation}, 
volume = {Proceedings of the International Astronautical Congress, IAC 2022}, 
publisher = {International Astronautical Federation, IAF}, 
year = {2022}, }

@Proceedings{ulrich,
author = {Ulrich, Steve}, 
title = {Nonlinear passivity-based adaptive control of spacecraft formation flying}, 
booktitle = {Nonlinear passivity-based adaptive control of spacecraft formation flying}, 
volume = {2016 American Control Conference (ACC)}, 
publisher = {IEEE}, 
pages = {7432-7437}, 
year = {2016}, }

@Article{tiwari,
author = {Tiwari, Madhur and Prazenica, Richard and Henderson, Troy}, 
title = {Direct adaptive control of spacecraft near asteroids}, 
journal = {Acta Astronautica}, 
volume = {202}, 
pages = {197–213}, 
year = {2023}, }

@Article{liu,
author = {Liu, Miao and Chowdhary, Girish and Castra da Silva, Bruno and Liu, Shih-Yuan and How, Jonathan P}, 
title = {Gaussian Processes for Learning and Control: A Tutorial with Examples}, 
journal = {IEEE Control Syst.}, 
volume = {38}, 
number = {5}, 
pages = {53–86}, 
year = {2018},}

@Article{chowdhary,
author = {Chowdhary, G and Kingravi, HA and How, JP and Vela, PA}, 
title = {Bayesian nonparametric adaptive control using Gaussian processes.}, 
journal = {IEEE Trans Neural Netw Learn Syst}, 
volume = {26}, 
number = {3}, 
pages = {537–550}, 
year = {2015}, }

@Book{alfriend2009,
author = {Alfriend, Kyle T and Vadali, Srinivas R and Gurfil, Pini and How, Jonathan P and Breger, Louis}, 
title = {Spacecraft formation flying: Dynamics, control and navigation}, 
volume = {2}, 
publisher = {Elsevier}, 
year = {2009}, }

@Book{aastrom2013adaptive,
  title={Adaptive control},
  author={{\AA}str{\"o}m, Karl J and Wittenmark, Bj{\"o}rn},
  year={2013},
  publisher={Courier Corporation}, }

@article{csato2002sparse,
  title={Sparse on-line Gaussian processes},
  author={Csat{\'o}, Lehel and Opper, Manfred},
  journal={Neural computation},
  volume={14},
  number={3},
  pages={641--668},
  year={2002},
  publisher={MIT Press One Rogers Street, Cambridge, MA 02142-1209, USA journals-info~…},}

@article{csato2000sparse,
  title={Sparse representation for Gaussian process models},
  author={Csat{\'o}, Lehel and Opper, Manfred},
  journal={Advances in neural information processing systems},
  volume={13},
  year={2000},}

@inproceedings{bowman2008new,
  title={A new empirical thermospheric density model JB2008 using new solar and geomagnetic indices},
  author={Bowman, Bruce and Tobiska, W Kent and Marcos, Frank and Huang, Cheryl and Lin, Chin and Burke, William},
  booktitle={AIAA/AAS astrodynamics specialist conference and exhibit},
  pages={6438},
  year={2008},}

@article{ono2021gnc,
  title={GNC design and evaluation of Hayabusa2 descent operations},
  author={Ono, Go and Terui, Fuyuto and Ogawa, Naoko and Mimasu, Yuya and Yoshikawa, Kent and Yasuda, Seiji and Matsushima, Kota and Takei, Yuto and Saiki, Takanao and Tsuda, Yuichi},
  journal={TRANSACTIONS OF THE JAPAN SOCIETY FOR AERONAUTICAL AND SPACE SCIENCES, AEROSPACE TECHNOLOGY JAPAN},
  volume={19},
  number={2},
  pages={259--265},
  year={2021},
  publisher={THE JAPAN SOCIETY FOR AERONAUTICAL AND SPACE SCIENCES},}





\end{document}